\documentclass[a4paper]{jpconf}
\usepackage{graphicx}
\usepackage{epsfig,graphics}

\begin{document}

\def\Ngr{\mbox{$\rm N_{\rm gr}$}}
\def\Ie{\mbox{$I_{\rm eff}$}}
\newcommand{\mie}{$<\! \mu\! >_{\rm e}$}

\def\eSF{\mbox{$\epsilon_{\rm SF}$}}
\def\Re{\mbox{$R_{\rm eff}$}}
\def\sig{\mbox{$\sigma$}}
\def\sigc{\mbox{$\sigma_0$}}
\def\sigro{\mbox{$\sigma_{0, \rm rot}$}}
\def\Msun{\mbox{$M_\odot$}}
\def\Lsun{\mbox{$L_\odot$}}
\def\Ysun{\mbox{$\Upsilon_\odot$}}
\def\ML{\mbox{$M/L$}}
\def\Yst{\mbox{$\Upsilon_{*}$}}
\def\YstB{\mbox{$\Upsilon_{*,B}$}}
\def\Ydyn{\mbox{$\Upsilon_{\rm dyn}$}}
\def\Ytot{\mbox{$\Upsilon_{\rm tot}$}}
\def\Mdyn{\mbox{$M_{\rm dyn}$}}
\def\mst{\mbox{$M_{*}$}}
\def\age{\mbox{$t_{\rm gal}$}}
\def\Zsun{\mbox{$Z_{\odot}$}}
\def\LB{\mbox{$L_{B}$}}
\def\MB{\mbox{$M_{B}$}}
\def\vmax{\mbox{$V_{\rm max}$}}
\def\lsim{\mathrel{\rlap{\lower3.5pt\hbox{\hskip0.5pt$\sim$}}
    \raise0.5pt\hbox{$<$}}}                
\def\gsim{~\rlap{$>$}{\lower 1.0ex\hbox{$\sim$}}}
\def\fDM{\mbox{$f_{\rm DM}$}}
\def\kms{\mbox{\,km~s$^{-1}$}}
\def\rhoDM{\mbox{$\langle \rho_{\rm DM} \rangle$}}

\title{Dark matter and alternative recipes for the missing mass}

\author{Crescenzo Tortora$^1$, Philippe Jetzer$^1$ and  Nicola R. Napolitano$^2$}

\address{$^1$Universit$\rm \ddot{a}$t Z$\rm \ddot{u}$rich, Institut f$\rm \ddot{u}$r
Theoretische Physik, Winterthurerstrasse 190, CH-8057, Z$\rm
\ddot{u}$rich, Switzerland }

\address{$^2$INAF -- Osservatorio Astronomico di Capodimonte, Salita
Moiariello, 16, 80131 - Napoli, Italy}

\ead{ctortora@physik.uzh.ch}

\address{}

\begin{abstract}
Within the standard cosmological scenario the Universe is found to
be filled by obscure components (dark matter and dark energy) for
$\sim 95\%$ of its energy budget. In particular, almost all the
matter content in the Universe is given by dark matter, which
dominates the mass budget and drives the dynamics of galaxies and
clusters of galaxies. Unfortunately, dark matter and dark energy
have not been detected and no direct or indirected observations
have allowed to prove their existence and amount. For this reason,
some authors have suggested that a modification of Einstein
Relativity or the change of the Newton's dynamics law (within a
relativistic and classical framework, respectively) could allow to
replace these unobserved components. We will start discussing the
role of dark matter in the early-type galaxies, mainly in their
central regions, investigating how its content changes as a
function of the mass and the size of each galaxy and few
considerations about the stellar Initial mass function have been
made. In the second part of the paper we have described, as
examples, some ways to overcome the dark matter hypothesis, by
fitting to the observations the modified dynamics coming out from
general relativistic extended theories and the MOdyfied Newtonian
dynamics (MOND).
\end{abstract}

\section{Introduction}

Within the standard cosmological framework, the so called
$\Lambda$ cold dark matter ($\Lambda$CDM) scenario, the Universe
is predicted to be filled by a huge amount ($\sim 95\%$) of
obscure energy components (dark matter, DM, and dark energy, DE),
and only $\sim 5\%$ is made by baryons (i.e. stars and gas).
According to several observations (i.e. distance moduli of high-z
Supernovae Ia, anisotropies in the cosmic microwave background
radiation, etc.) the Universe is spatially flat and in an
accelerated phase of expansion. DE amounts to $\sim 75\%$ of the
total energy budget, is modelled as a negative equation of state
and is the main driver of the observed cosmic acceleration
\cite{perl, reiss, spe07}. DM has an important role not only at
the cosmic scale but also to explain the growth of the structures
and the evolution of galaxies across the cosmic time. Although
this is the widest accepted theory to explain the various
astronomical and cosmological observations, there has been no
direct observational evidence for both DM and DE. Anyway, this has
motivated the investigation of alternative theories to explain
observations without invoking the presence of such unknown
ingredients. We will first concentrate on DM in galaxies (mainly
massive ellipticals) and on some alternative recipe. In particular
we will briefly discuss the viability of two approaches: 1) a
relativistic one consisting in a generalization of the Einstein'
equations, the so called $f(R)$ theories, and 2) a classical one,
by means of a modification of the second Newton's law, as the
MOdified Newtonian Dynamics (MOND).


\section{The missing mass}\label{sec:missing_mass}

The need of DM in galaxies and clusters was firstly suggested by
Fritz Zwicky in 1934 to account for evidence of  ``missing mass''
in the observed orbital velocities of galaxies in clusters
\cite{Zwicky}. Applying the virial theorem to the Coma cluster he
found a total mass 400 times larger than that visually observed.
Till 60s, 70s no other observational evidences for the existence
of DM were found. However, it was only later, in 70s, that the
evidences started to be clear enough to put the DM at the center
of the astronomical community attention. In these years, some
authors observed that the stars and gas in spiral galaxies have an
orbital velocity which remains constant even beyond their optical
radii (i.e., beyond the region where much of the stars are
located) \cite{Rubin_Ford70, Bosma81}. Nowadays, many observations
have pointed out the existence of this missing mass problem in
many spirals and other kinds of galaxies, as low surface
brightness and ellipticals, and clusters \cite{deBlok+01,
Salucci_Borriello, Bradac+08}. The mass distribution within these
different astrophysical objects is investigated by means of
different mass probes. In particular, rather than circular
velocity in spirals, strong (and weak) gravitational lensing,
velocity dispersion measurements and X-ray observations are
powerful mass tracers inside ellipticals and clusters of galaxies.

Lately, important progresses have been made also in the study of
the mass distribution in elliptical galaxies, mainly thanks to the
development of the discrete velocity mapping with globular
clusters (GCs; \cite{Romanowsky+09}) and in particular the doppler
measurement of the OIII emission from Planetary Nebulae
\cite{Napolitano+01,Napolitano+02,Romanowsky+03, DeLorenzi+09,
Napolitano+09, Napolitano+11}. PNe in particular (being stars)
have allowed the stellar kinematics in the (otherwise
inaccessible) outer regions of such galaxies to be studied
\cite{Coccato+09}.

\section{Dark matter}\label{sec:DM}

Within the Einstein framework, the missing mass needs to be filled
by DM, which seems to be dominant in the very external regions of
galaxies, where the stars are absent, and also in the central
regions, depending on the stellar ingredients which have been
adopted. Although in the central regions many uncertainties in the
DM content determination arise from the contribution of stars,
many kinds of observations probing the mass in these regions and
for huge samples of galaxies are available. In Tortora et al. 2009
\cite{Tortora2009} and Napolitano, Romanowsky \& Tortora 2010
\cite{NRT10}, we have recently investigated the DM content within
1 \Re\ (i.e., the radius where is enclosed one-half of the total
projected light in the galaxy) for a sample of local ellipticals
from the sample in Prugniel \& Simien 1996 \cite{PS96}. The
stellar mass, \mst, is determined by fitting synthetic spectral
models to the measured colours, while dynamical (total) mass,
\Mdyn\ within 1 \Re, is recovered by adopting a singular
isothermal sphere (SIS) for the total mass
distribution\footnote{An isothermal profile has been shown to
reproduce quite well  the total mass profile in massive
ellipticals \cite{SLACSIII,SLACSIV}} and the observed central
velocity dispersion is used to infer the best mass profile (see
\cite{Tortora2009} for further details).

Djorgovski \& Davis 1987 \cite{DD87} and Dressler et al. 1987
\cite{Dressler87} discovered a tight relation among central
velocity dispersion \sigc, the effective radius \Re, and effective
surface brightness \Ie, the so-called Fundamental Plane (FP). It
has been observed a deviation of its coefficients from the virial
theorem expectation under the assumptions of  homology and
constant mass-to-light ratio  (\ML). One  possible explanation of
the tilt is a variation of the total \ML \, with galaxy
luminosity/mass \cite{Dressler87}, often parameterized as a power
law,  $\ML \propto  L^{\gamma}$. This relation  might reflect a
variation of different galaxy properties with mass. Thus, we have
evaluated the impact  on $\gamma$ of various factors, like stellar
population properties (metallicity, age and SF history), IMF,
rotational support, luminosity profile, non-homology and DM
fraction. We found that the stellar \ML\ contributes little to the
tilt, suggesting that the tilt is ascribed to DM.


\begin{figure}
\psfig{file=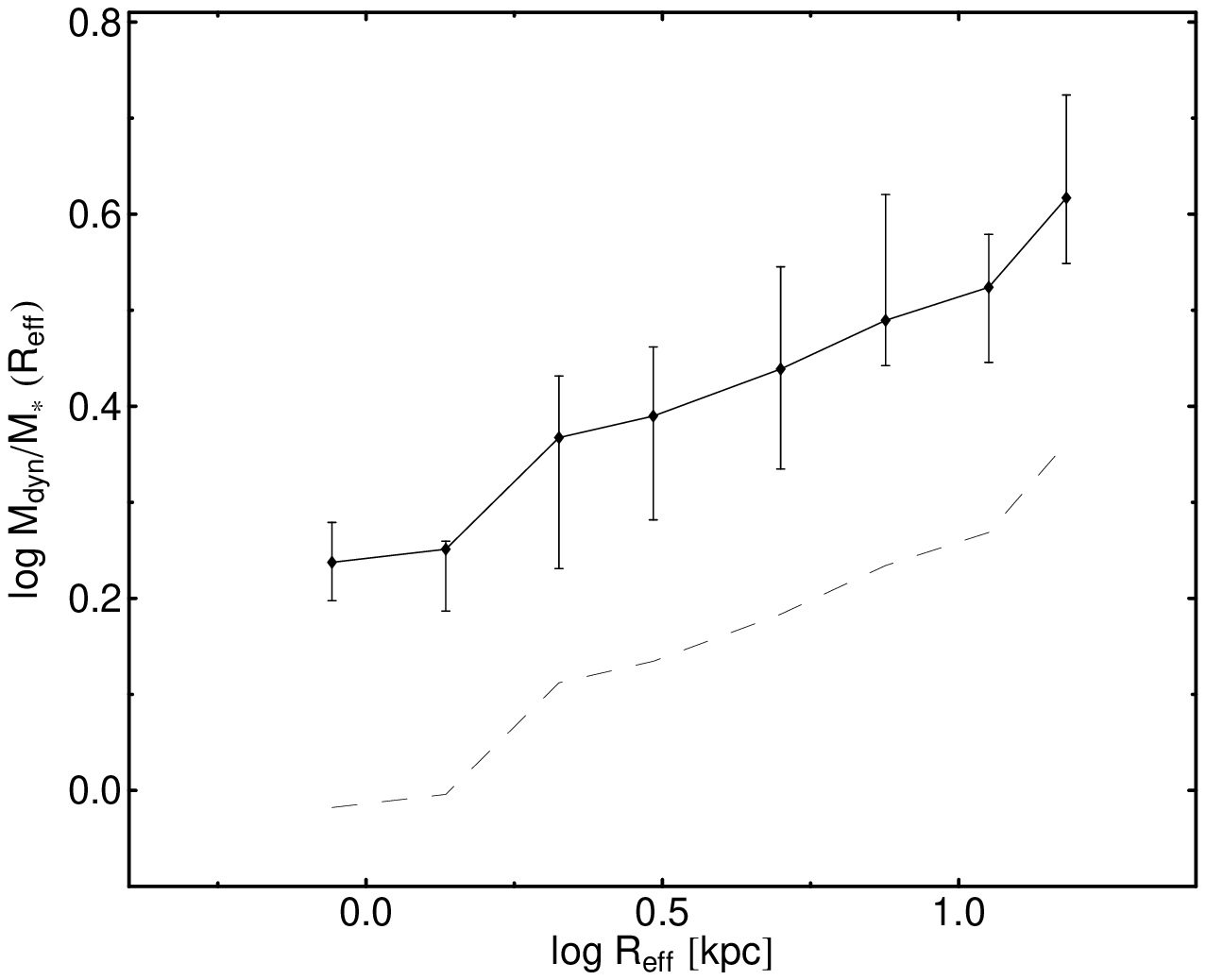,
width=0.47\textwidth}\psfig{file=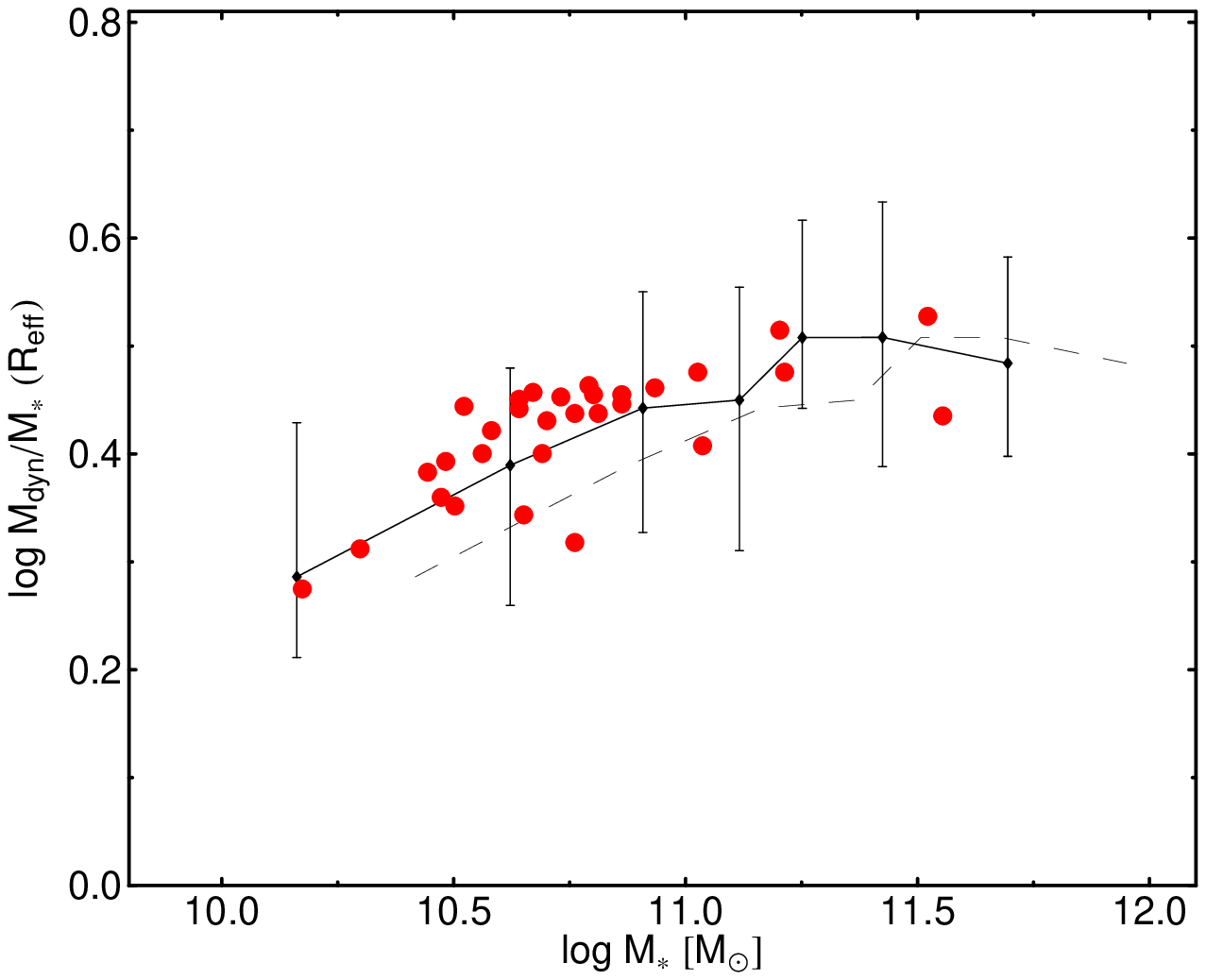,
width=0.47\textwidth}\caption{Dynamical-to-stellar mass ratio as a
function of \Re\ (left) and \mst\ (right) for the pure ellipticals
from Prugniel \& Simien 1996 \cite{PS96}. Continue lines and bars
are for medians and 25-75th quantiles when a Chabrier IMF is
adopted, while the dashed lines are the median trends for a
Salpeter IMF. The red points in the right plot are the results for
simulated central galaxies in Ruszkowski \& Springel 2009
\cite{RS09}.} \label{fig:fig1}
\end{figure}

We surveyed trends of the DM fraction within 1 \Re, finding it to
be increasing with luminosity and stellar mass. Our conclusions
are sensitive to various systematic uncertainties which we have
investigated in detail, but are consistent with the results of
dynamics studies at larger radii \cite{Tortora2009}. We also find
that stronger correlations are found in terms of central velocity
dispersion and effective radii, which means that at larger size,
we go deeper in the halo, finding more DM dominated regions
\cite{NRT10}. In particular, in Fig. \ref{fig:fig1} we show the
deprojected dynamical-to-stellar ratio (adopting a Chabrier IMF
\cite{Chabrier01}) as a function of \Re\ and \mst\ for the pure
elliptical galaxies in our sample. We find that $\Mdyn/\mst
\propto \Re^{0.27}$ and $\Mdyn/\mst \propto \mst^{0.13}$, this
last being steeper than the slope of $\sim 0.06$ found in Hyde \&
Bernardi 2009 \cite{HB09a}. In the smallest and less massive
galaxies is $\Mdyn/\mst \sim 1.5$, i.e. DM is $\sim 40\%$ of the
total mass, while in the largest and most massive galaxies is
$\Mdyn/\mst \sim 3-4$, corresponding to a DM fraction of $\sim
70\%$. A Salpeter IMF \cite{Salpeter55} produces larger stellar
masses, and thus lower $\Mdyn / \mst$. A rigid shift is observed
in the $\Mdyn / \mst - \Re$ plot, null DM fractions are found in
the smaller galaxies and $\sim 50\%$ of DM in the largest systems.
%
%
We also compare our results with the simulated galaxies in
Ruszkowski \& Springel 2009 \cite{RS09}, finding a very good
agreement in the average values, but the scatter in the simulated
data is smaller. We have confirmed the increasing trend with mass
and \Re\ adopting a sample of intermediate redshift lens galaxies
from SLACS survey \cite{Auger+09} with available lensing and
dynamical information, varying the stellar recipe and the galaxy
model in \cite{Cardone+09, CT10, Cardone+11}.

In \cite{NRT10} we have also examined the correlations between
masses, sizes and SF histories. We confirmed an anticorrelation
between \Re\ and stellar age and went on to survey for trends with
the central content of DM. An average relation between the central
DM density and galaxy size of $<\rho_{\rm DM}> \propto \Re^{-2}$
provides the first clear indication of cuspy DM haloes in these
galaxies. From the comparison with $\Lambda$CDM expectations (see
details in \cite{NRT10, Tortora10lensing}), our data are found to
be consistent with a Chabrier or Kroupa IMF if the halo is
adiabatically contracted\footnote{Within $\Lambda$CDM, (DM only)
N-body simulations predict that galaxy and galaxy cluster halos
follow a universal mass density profile, the so called Navarro,
Frenk \& White (NFW) profile, \cite{NFW96}. While gas is falling
down in the DM potential well, stars form and drag in the centers
some DM, adiabatically contracting the DM distribution
\cite{Gnedin+04}.}, while, if we want to retain the original
outcomes from N-body simulations (i.e. the NFW profile), a
Salpeter IMF is best suited (in agreement with \cite{Treu+10}).
The DM density scales with galaxy mass as expected, deviating from
a universal halo profile for dwarf and spiral galaxies (see also
\cite{CT10, Cardone+11}).

The IMF inferred from star counts in local spiral galaxies is
bottom-light (as a Chabrier or Kroupa \cite{Kroupa01}), but no
direct observations are possible for other galaxy types (as, e.g.,
ellipticals), different environments and at higher redshift.
Therefore, the IMF is a fundamental ingredient which has still to
be fully understood. Anyway, Chabrier and Salpeter IMFs are
assumed as the limiting cases, producing a difference in stellar
masses of $0.25 \, \rm dex$ and we have shown, that also assuming
a Salpeter IMF a residual DM fraction is surviving. In order to
have, on average, $\Mdyn = \mst$ we need a IMF producing stellar
M/L ratios $\sim 2.7_{-0.9}^{+1.4}$ ($1 \, \sigma$ uncertainty)
times the ones from a Chabrier IMF. For only $\sim 20\%$ of the
galaxies in our sample a Salpeter IMF is totally filling the gap,
avoiding the need of DM (mainly in the smallest galaxies, see left
panel in Fig. \ref{fig:fig1}), but for many of the galaxies larger
stellar M/L would be needed. Such very high stellar masses could
be produced by 1) a power-law IMF with a slope steeper than the
Salpeter one\footnote{In Tortora et al. 2009 \cite{Tortora2009} we
have shown that if a power-law IMF with a slope of $x=-1.85$ is
adopted, then the stellar M/L ratio are $\sim 3.2$ times the ones
from a Chabrier IMF.} \cite{BdeJ2001} and equivalently 2) a
``bottom-heavier'' and/or a ``top-lighter'' IMF. But, no
indication exists in favour of such IMF alternatives in the local
universe and they seem quite unrealistic, suggesting that DM is
still an unavoidable ingredient. Anyway, recently van Dokkum \&
Conroy \cite{vDC10}, analyzing the spectra of massive ellipticals
in the Virgo and Coma clusters, have found that low mass stars are
very abundant (contributing to $\sim 60\%$ of the total stellar
mass) in such galaxies and point to a steeper IMF, which in
according to the comments above, would fill the gap, without
needing DM, at least in the central regions.

Finally, we introduced a new fundamental constraint on galaxy
formation by finding that the central DM fraction decreases with
stellar age. This result is only partially explained by the
size-age dependencies, and the residual trend goes on the opposite
direction to basic DM halo expectations. Therefore, we suggested
that there may be a connection between age and halo contraction or
IMF and that galaxies forming earlier had stronger baryonic
feedback, which expanded their haloes, or lumpier baryonic
accretion, which avoided halo contraction, or a lighter IMF. Using
the intermediate-redshift galaxy sample from the SLACS survey, we
have a confirmation of such results and a negligible evidence of
galaxy evolution over the last $\sim 2.5$ Gyr other than passive
stellar aging \cite{Tortora10lensing}.

\section{MOdified Newtonian gravity}\label{sec:MOND}

The results discussed above have been obtained assuming that the
Einstein relativistic theory, and the classical Newtonian theory
of gravity as its limit, may be used also on galactic scales.
However, the outer regions of galaxies typically are in a low
acceleration regime, and in this regime Newtonian dynamics has
never been experimentally tested. Motivated by this consideration,
Milgrom (1983) \cite{Milgrom83} proposed to modify Newton's second
law of dynamics as $F = mg$, where the acceleration $g$ is now
related to the Newtonian one $g_{\rm N}$ as $g \, \mu(g/a_{0}) =
g_{\rm N}$.  The theory thus obtained is referred to as MOND.
Here, $a_0$ is a new universal constant and $\mu(x)$ may be an
arbitrary function with the properties $\mu (x>>1) = 1$ and $\mu
(x << 1) = x$, i.e. Newton's law is recovered in high acceleration
regimes, while at extremely low accelerations we have the
deep-MOND regime, i.e. $F \propto g^{2}$. Since at large distance
from the center is $\mu(g/a_{0}) = g/a_{0}$, the relation $m
\mu(g/a_{0}) g = GMm/r^{2}$ becomes $g^{2}/a_{0} = GM/r^{2}$ or $g
= \sqrt{G M a_{0}}/r$. Using the relation between acceleration and
velocity in a circular orbit we find $v = \sqrt[4]{G M a_{0}}$.
Thus, MOND reproduces observed rotation curves in spiral galaxies
(e.g. \cite{McGaugh05,McGaugh08}), gives a theoretical
interpretation of the empirically determined Tully-Fisher law
\cite{McGaugh05}, and also works in dwarf spheroidals
\cite{Angus08}. On the contrary, clusters seem to need extra DM to
accommodate MOND \cite{PS05}, although recently it has been shown
that sterile neutrinos can fill the gap \cite{Angus09}.

We have considered a sample of $\sim 9000$ local ellipticals from
the SDSS sample \cite{Blanton+05} and a Chabrier IMF is adopted.
For the first time we have shown that MOND is able to reproduce
the dynamics in the central regions (typically $\lsim \Re$) of a
large sample of local ellipticals. The boost to the velocity
dispersion in the MOND scenario helps to reduce the need of both
radial anisotropy (in the Newtonian case without DM), deviations
from a Chabrier IMF or DM. We also showed that MOND is able to
predict a FP for ellipticals (similarly to the Newtonian case),
but a tilt between the observed and the MOND FP is found
\cite{Cardone+11MOND}. We are working to study if MOND is able to
reproduce the dynamics of both ellipticals and spirals,
investigating more carefully the role of anisotropy and of the
IMF.

\section{Extended theories of gravity}\label{sec:fR}

An alternative approach to the missing mass problem consists to
modify the Einstein's equations, 1) or by introducing a field
(with non zero mass) which is coupled with matter, as in the so
called tensor-vector-scalar (TeVeS) theories
\cite{Moffat06,BM07,MT09}, or 2) by replacing the Ricci tensor $R$
in the Einstein-Hilbert action with a function of R, $f(R)$.
Although the TeVeS theories have been shown to be successful to
reproduce different observations (e.g. \cite{BM07}), we will
concentrate on the $f(R)$ theories, which have been suggested to
reproduce both DE on cosmic scales \cite{CCCT03} and DM
\cite{SM02,CPRS09} (see also
\cite{CdL11,Clifton+11,Nojiri_Odintsov11} for a complete review on
the subject).

It has been demonstrated that very general $f(R)$ analytical
theories can induce a modification in the dynamics of massive
particles. In particular, if one solves the field equations in the
weak field limit under the general assumption of an analytic
Taylor expandable $f(R)$ functions of the form
\begin{equation}
f(R) = f_{0} + f_{1} R + f_{2} R^{2} + .....
\end{equation}
the new gravitational potential can be written as
\begin{equation}
\phi(r)= - \frac{GM}{(1+\delta)r} \left (1+ \delta
e^{-\frac{r}{L}} \right ) \label{eq:yukawa}
\end{equation}
where the first term is the Newtonian-like part of the potential
associated to baryonic point-like mass $M/(1+ \delta)$ (no DM) and
the second term, the Yukawa-like potential, is a modification of
the gravity including a scale length, L, associated to the above
coefficients of the Taylor expansion. This gravitational potential
was also adopted by Sanders 1984 \cite{Sanders84}, under the
assumption of anti-gravity generated by massive particles (the so
called Finite Length-scale Anti-Gravity, FLAG, theory). This
theory conjectures that in addition to the massless graviton,
particles with mass can carry the gravitational force, an
anti-gravity force (if $-1 < \delta < 0$), which can allow to
solve the missing mass problem. We have performed the first
analysis of extended stellar kinematics (up to $7$ \Re s) of 3
ellipticals (NGC 3379,  NGC 4374,  NGC 4494), where the
Yukawa-like gravitational potential in Eq. (\ref{eq:yukawa}) is
considered \cite{N+11_fR}. We find that these modified potentials
are able to fit quite well all galaxies in our sample and the
orbital anisotropy distribution turns out to be similar to the one
estimated if a dark halo is considered. The parameter which
measures the strength of the Yukawa-like correction, $\delta$
($=-0.88, -0.79, -0.75$ for the three galaxies) is, on average,
larger than the one found previously in spiral galaxies
(\cite{Sanders84}, $-0.95 \lsim \delta \lsim -0.92$) and seemingly
correlating with the orbital anisotropy.

Though the additional Yukawa term in the gravitational potential
modifies dynamics with respect to the standard Newtonian limit of
General Relativity, we have demonstrated that the motion of
massless particles results unaffected thanks to suitable
cancellations in the post-Newtonian limit \cite{Lubini+11}. Thus,
all the lensing observables are equal to the ones known from
General Relativity. We are planning to test the implications of
these results by fitting both lensing and dynamical observables in
galaxy and cluster lensing events. In particular, following
\cite{Tortora10lensing}, the sample of lens galaxies at
intermediate  redshift from SLACS sample will be an efficient
testing ground for these extended theories.

\section{Conclusions}\label{sec:conclusions}

In this contribution we have discussed the missing mass problem,
which come out from different observations in galaxies and
clusters. In a standard Newtonian and Einstein approach, DM is an
unavoidable ingredients to explain both cosmological observations
and dynamics in galaxies and clusters. DM is found to be the
dominant component, at all, since the stellar content is
negligible with respect to the total mass determined by, e.g.
dynamical analysis. Less obvious is the DM role and its amount in
the central regions (typically $\lsim \Re$) where a very large
fraction of stars are settled. We have shown that DM in the
central regions is strongly dependent on the IMF assumption. In
fact, while the typical ``bottom-light'' (Chabrier or Kroupa)
IMFs, found through direct observations in local spirals, predict
lower stellar masses and thus larger DM fractions, $\sim 40-70\%$
in massive ellipticals (with stellar mass $\mst \gsim 10^{10}
\Msun$), a Salpeter IMF predict larger stellar masses, and thus
lower DM fractions $\sim 0-50\%$. We have also shown how the
typical halo density profiles derived from $\Lambda$CDM N-body
simulations are consistent with the only if stars are distributed
in according to a Salpeter IMF, while a contraction of the DM halo
is needed to reconcile a Chabrier IMF. Moreover, to avoid any DM
in the central regions, we would need steeper IMFs, but no strong
observation seems now convincing us which this is the case (e.g.
\cite{vDC10}). The situation in the external regions (few \Re s up
to the virial radius) is more complicated, and such a steep IMF
seems to us not enough to replace the amount of missing mass
needing in such external regions.

Anyway, until no direct indication of the existence of DM will be
provided by the many working (or future) experiments, people are
encouraged to search for some alternative theories to overcome the
need of such unseen matter source. This can be done by means of
different approaches, within both a relativistic theory, as an
extension of the General Relativity, and within a classical
Newtonian approach, like in the MOND. We have shown as both MOND
or $f(R)$ theories are able to reproduce a wide set of
observations, in particular, the dynamics in different kinds of
galaxies, and recently both central ($\lsim \Re$) and more
external (few \Re) dynamical data in ellipticals
\cite{Cardone+11MOND,N+11_fR}.

In the future we expect to investigate the correlations between DM
content, mass, IMF and stellar population parameters, checking
what is the impact on such correlations when the gravity theory is
changed. Such theories need to be tested on wider samples of
(different type) galaxies, and galaxy clusters, relying on
different kinds of observations (gravitational lensing, dynamics,
X-ray, etc.), in order to check if DM can be avoided at all,
everywhere in galaxies or if our constrain on DM in galaxies have
to be revised downward. To refuse/validate any of such theories,
as alternatives to DM, we need a) to fit data quite good, at least
as much as in the DM case \cite{N+11_fR}, b) that some parameters
derived from the fitting procedure (as other model parameters,
velocity dispersion anisotropy, stellar mass-to-light ratio if
allowed to be free, etc.) assume physically reasonable values
\cite{Cardone+11MOND,N+11_fR} and c) different observational
probes give consistent results. If such points would be addressed,
then the theory analyzed can be considered as a viable
alternative. Anyway, a definite answer about the existence of DM
and its abundance will arrive from both direct and indirect
experiments in progress and scheduled to start in the few next
years.


\section*{References}
\begin{thebibliography}{50}

\bibitem{perl} Perlmutter S. et al. 1999 {\it ApJ} {\bf 517} 565

\bibitem{reiss}  Reiss A.G. et al. 1998 {\it AJ} {\bf 116} 1009

\bibitem{spe07} Spergel D.N. et al. 2007 {\it ApJS} {\bf 170} 377


\bibitem{Zwicky} Zwicky, F. 1937 {\it ApJ} {\bf 86} 217

\bibitem{Rubin_Ford70}  Rubin V. C. \& Ford W. K. Jr. 1970 {\it ApJ} {\bf 159} 37

\bibitem{Bosma81} Bosma A. 1981 {\it AJ} {\bf 86} 1825

\bibitem{deBlok+01} de Blok W. J. G., McGaugh S. S., Bosma A. \& Rubin V. C. 2001 {\it ApJ} {\bf 552} 23

\bibitem{Salucci_Borriello} Salucci P. \& Borriello A.\ 2003 {\it Particle Physics in the New Millennium} {\bf 616} 66


\bibitem{Bradac+08} Brada$\rm \check{c}$ M. et al. 2008 {\it ApJ} {\bf 687} 959


\bibitem{Romanowsky+09} Romanowsky A. J., Strader J., Spitler L. R., Johnson R.,
Brodie J. P., Forbes D. A. \& Ponman T. 2009 {\it AJ} {\bf 137}
4956

\bibitem{Napolitano+01} Napolitano, N.~R.,
Arnaboldi, M., Freeman, K.~C., \&   Capaccioli, M.\ 2001 {\it
A\&A} {\bf 377} 784

\bibitem{Napolitano+02} Napolitano, N.~R., Arnaboldi,
M., \&  Capaccioli, M.\ 2002, {\it A\&A} {\bf 383} 791

\bibitem{Romanowsky+03} Romanowsky A. J., Douglas N. G., Arnaboldi M., Kuijken K.,
Merrifield M. R., Napolitano N. R., Capaccioli M. \& Freeman, K.
C. 2003 {\it Science} {\bf 301} 1696

\bibitem{DeLorenzi+09} De Lorenzi F. et al. 2009 {\it MNRAS} {\bf 395} 76

\bibitem{Napolitano+09} Napolitano N. R. et al. 2009  {\it MNRAS} {\bf 393} 329

\bibitem{Napolitano+11} Napolitano N. R. et al. 2011 {\it MNRAS} {\bf 411} 2035



\bibitem{Coccato+09} Coccato L. et al. 2009  {\it MNRAS} {\bf 394} 1249


\bibitem{PS96} Prugniel Ph. \& Simien F. 1996 {\it A\&A} {\bf 309}
749

\bibitem{Tortora2009} Tortora C. et al., 2009 {\it MNRAS} {\bf 396} 1132

\bibitem{NRT10} Napolitano N. R., Romanowsky A.J. \& Tortora C., 2010 {\it  MNRAS} {\bf 405} 2351

\bibitem{DD87} Djorgovski S.  \&  Davis M.  1987 {\it ApJ} {\bf 313} 59

\bibitem{Dressler87} Dressler A., Lynden-Bell D., Burstein D., Davies R.~L.,  Faber S.~M., Terlevich R.,
Wegner G.,  1987 {\it ApJ} {\bf 313} 42

\bibitem{Cardone+09} Cardone V. F., Tortora C., Molinaro R., Salzano V., 2009 {\it A\&A} {\bf 504} 769

\bibitem{CT10} Cardone V.F. \& Tortora C., 2010 {\it MNRAS} {\bf 409} 1570

\bibitem{Cardone+11} Cardone V. F., Del Popolo A., Tortora C., Napolitano N.R. 2011 {\it MNRAS} {\bf 416} 1822

\bibitem{HB09a} Hyde J. B. \& Bernardi M. 2009 {\it MNRAS} {\bf 394} 1978

\bibitem{Tortora10lensing} Tortora C., Napolitano N.R., Romanowsky A.J., Jetzer Ph. 2010 {\it ApJ} {\bf 721} 1


\bibitem{SLACSIII} Koopmans L. V. E., Treu T., Bolton A.
S., Burles S., Moustakas L. A. 2006 {\it ApJ} {\bf 649} 599

\bibitem{SLACSIV} Gavazzi R. et al. 2007  {\it ApJ} {\bf 667} 176


\bibitem{Auger+09} Auger M. W., Treu T., Bolton A. S., Gavazzi R.,
Koopmans L. V. E., Marshall P. J., Bundy K. \& Moustakas L. A.
2009  {\it ApJ} {\bf 705} 1099


\bibitem{Salpeter55} Salpeter E.E. 1955 {\it ApJ} {\bf 121}
161

\bibitem{Kroupa01} Kroupa P. 2001  {\it MNRAS} {\bf 322} 231

\bibitem{Chabrier01} Chabrier G. 2001  {\it ApJ} {\bf 554} 1274


\bibitem{RS09} Ruszkowski M. \& Springel V. 2009, ApJ, 696, 1094

\bibitem{NFW96}  Navarro J. F.,  Frenk C.F. \&  White S.D.M. 1996 {\it ApJ} {\bf 462} 563


\bibitem{Gnedin+04}  Gnedin O. Y., Kravtsov A. V., Klypin A. A. \&
Nagai D. 2004 {\it ApJ} {\bf 616} 16

\bibitem{Treu+10} Treu T., AugerM. W., Koopmans L. V. E., Gavazzi R., Marshall P. J.,
Bolton A. S. 2010 {\it ApJ} {\bf 709} 1195


\bibitem{BdeJ2001} Bell E. F. \& de Jong R.
S. 2001  {\it ApJ} {\bf 550} 212


\bibitem{vDC10} van Dokkum P. G. \& Conroy C. 2010 {\it Nature} {\bf 468} 940

\bibitem{Milgrom83} Milgrom M., 1983 {\it ApJ} {\bf 270} 365

\bibitem{McGaugh05} McGaugh S. S., 2005 {\it ApJ} {\bf 632} 859

\bibitem{McGaugh08} McGaugh S. S., 2008 {\it ApJ} {\bf 683} 137

\bibitem{Angus08} Angus G. W. 2008 {\it MNRAS} {\bf 387} 1481



\bibitem{PS05} Pointecouteau E. and Silk J. 2005 {\it MNRAS} {\bf 364} 654

\bibitem{Angus09} Angus G. W. 2009 {\it MNRAS} {\bf 394} 527


\bibitem{Blanton+05} Blanton M. R., Lupton R. H., Schlegel D. J., Strauss M. A.,
Brinkmann J., Fukugita M., Loveday J. 2005  {\it ApJ} {\bf 631}
208


\bibitem{Cardone+11MOND}  Cardone V.F., Angus G., Diaferio A., Tortora C., Molinaro R.  2011  {\it MNRAS} {\bf 412} 2617



\bibitem{Moffat06} Moffat J. W. 2006 {\it JCAP} {\bf 03} 004

\bibitem{BM07} Brownstein J. R., Moffat J. W. 2007 {\it MNRAS} {\bf 382} 29


\bibitem{MT09} Moffat J. W., Toth V. T. 2009 {\it MNRAS} {\bf 397} 1885


\bibitem{CCCT03} Capozziello S., Cardone V. F., Carloni S. and Troisi A.,
2003 {\it International Journal of Modern Physics} {\bf D12} 1969.

\bibitem{SM02} Sanders R.H. and  McGaugh S. S., 2002 {\it ARA\&A} {\bf 40} 263

\bibitem{CPRS09} Capozziello S., Piedipalumbo E., Rubano C. and Scudellaro P.,
2009 {\it A\&A} {\bf 505} 21

\bibitem{CdL11} Capozziello S., de Laurentis M. 2011 {\it PhR} {\bf 509} 167

\bibitem{Clifton+11} Clifton T. Ferreira, Pedro G., Padilla A., Skordis C. 2011,
arXiv:1106.2476

\bibitem{Nojiri_Odintsov11} Nojiri S., Odintsov S. D. 2011 {\it PhR} {\bf 505} 59

\bibitem{Sanders84} Sanders R. H. 1984 {\it A\&A} {\bf 136} L21

\bibitem{N+11_fR} Napolitano N.R., Capozziello S.,
Romanowsky A.J., Capaccioli M.  \& Tortora C. 2012, to appear in
ApJ, arXiv:1201.3363

\bibitem{Lubini+11} Lubini, M., Tortora, C., N\"af, J., Jetzer, Ph., Capozziello, S.
2011, {\it EPJC} {\bf 71} 1834, arXiv:1104.2851

\end{thebibliography}
\end{document}